# Optical teardown of a Kindle Paperwhite display by OCT

Bart Johnson, Walid Atia, Mark Kuznetsov, Noble Larson, Eric McKenzie, Vaibhav Mathur, Brian Goldberg, Peter Whitney

*Axsun Technologies, 1 Fortune Drive, Billerica, MA 01821, USA*

**Abstract:** An optical teardown, or reverse engineering, of an Amazon Kindle Paperwhite electrophoretic display was performed by Optical Coherence Tomography at 1060 nm. The display incorporates an optical diffuser, lightguide and scattering layers for white light illumination, capacitive touch sensing, and an electrophoretic display. All these layers can be imaged by OCT as well as the thin film transistor array on the back side for driving the pixels. Phase sensitive OCT is used to measure motion of the pigment particles as the display changes between black and white.

## 1. Introduction

The Amazon Kindle Paperwhite [1] is an advanced electronic book reader that features a black and white electrophoretic display [2] as well as capacitive multitouch screen capability and an internal lightguide with an optical scattering layer [3,4] to provide uniform, white light illumination. Teardowns, where a new electronic gadget is disassembled and photographed are commonly found on technically oriented web sites. Here we perform an "optical teardown" of the first generation Kindle Paperwhite display by Optical Coherence Tomography (OCT). This highlights an industrial application of OCT in the 1060 nm wavelength range using advanced swept laser sources and data acquisition hardware [5].

The heart of the Paperwhite is the electrophoretic display manufactured by E Ink Corporation [2]. Microencapsulated electrophoretic displays were first developed at MIT [6] and then refined [7] and commercialized [2]. The electrophoretic display layers consist of microcapsules of liquid containing black and white particles. The white particles are permanently charged negative and the black particles positive. They can thus be moved through electrophoresis by an electric field. Moving the white particles to the top of the capsule and black to the bottom results in a white pixel. Reversing the electric field turns it black. An interesting feature of these displays is that with no external electric field applied they remain in the same state. An image can be retained almost indefinitely with no power applied to the display. This fact, and the fact that the display emits no light, only reflects light, means that electrical power consumption is very low.

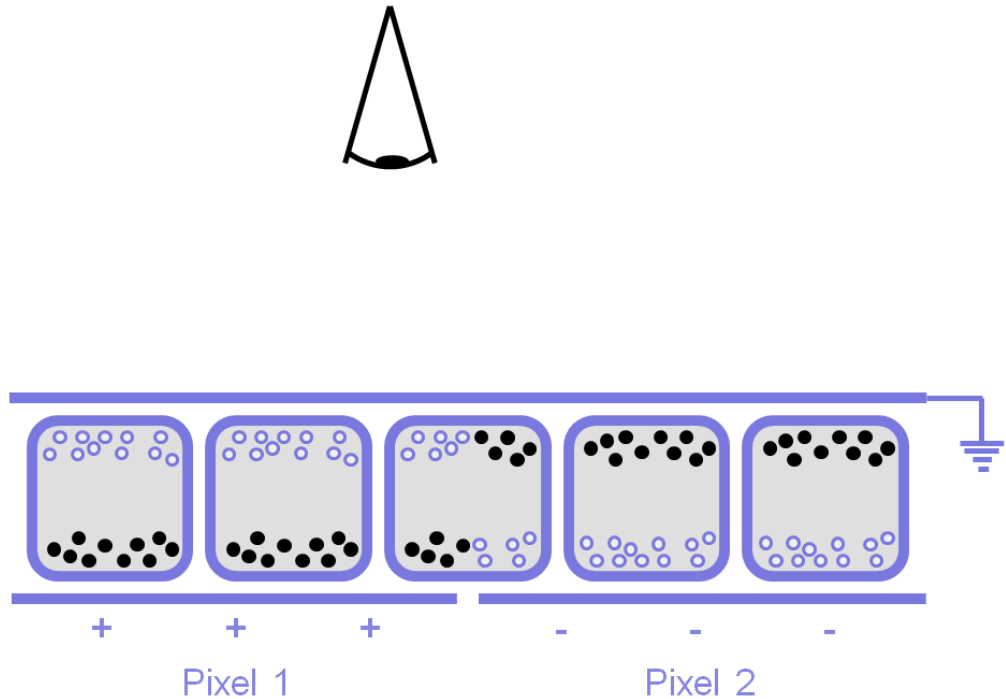

Figure 1.  Diagram of a microcapsule electrophoretic display.

2.  Overall display construction

The Kindle Paperwhite display is a complex layered device of glass, plastic and thin films, in addition to the electrophoretic pigment microcapsules.  A cross-section of the display taken with a 1060 nm OCT system is shown in Figure 2 and three dimensional renderings are shown in Figure 3.  The top layer is a light diffuser to remove reflected light glare.  It is followed by a high index light guide layer that conducts light from four white LEDs across the display.  A graded surface of microprinted scattering centers [3,4] directs white light out of the guide layer to uniformly illuminate the pigment layer.  Underneath the light guide is an 18 x 14 thin film capacitive touch screen layer. The capacitive array pitch is 7.0 x 6.6 mm on a 122 x 90 mm display area. The pigment layer contains fluid-filled microcapsules with charged black and white pigment particles.  The capsules are sandwiched between a transparent ground plane and thin film transistor (TFT) driven pixel electrodes unseen below the pigment layer.

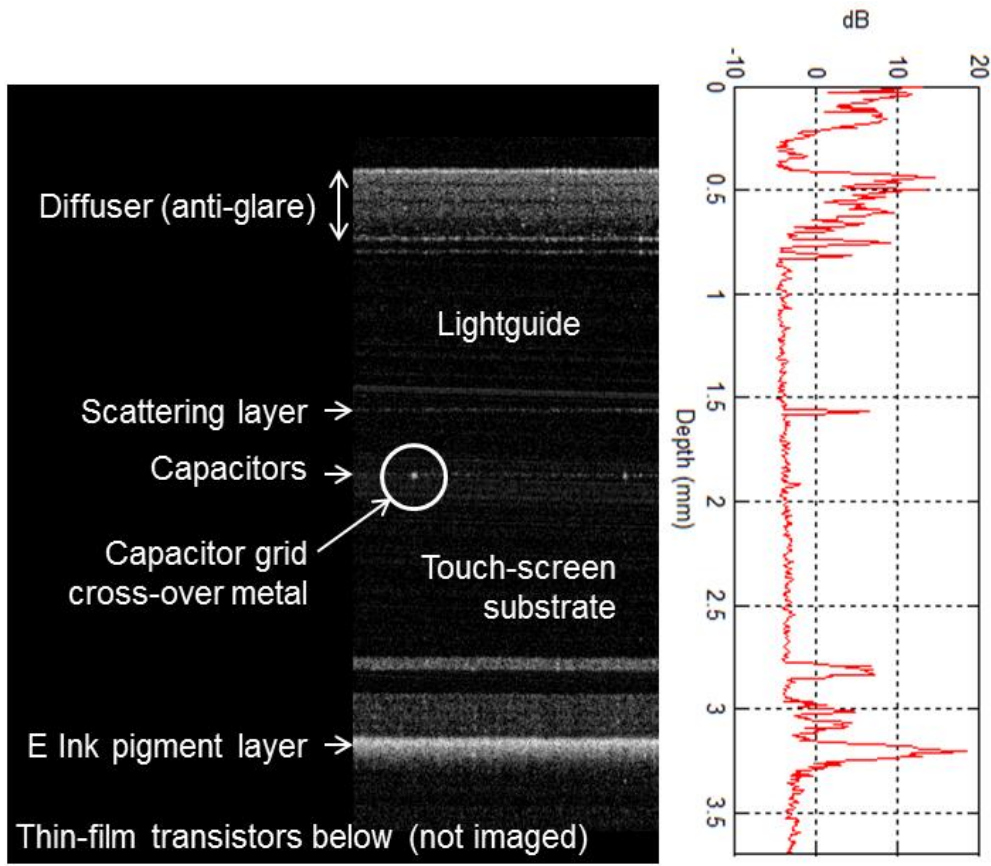

Figure 2.  Cross section image (left) and single A-line (right) of a Kindle Paperwhite display by OCT at 1060nm.

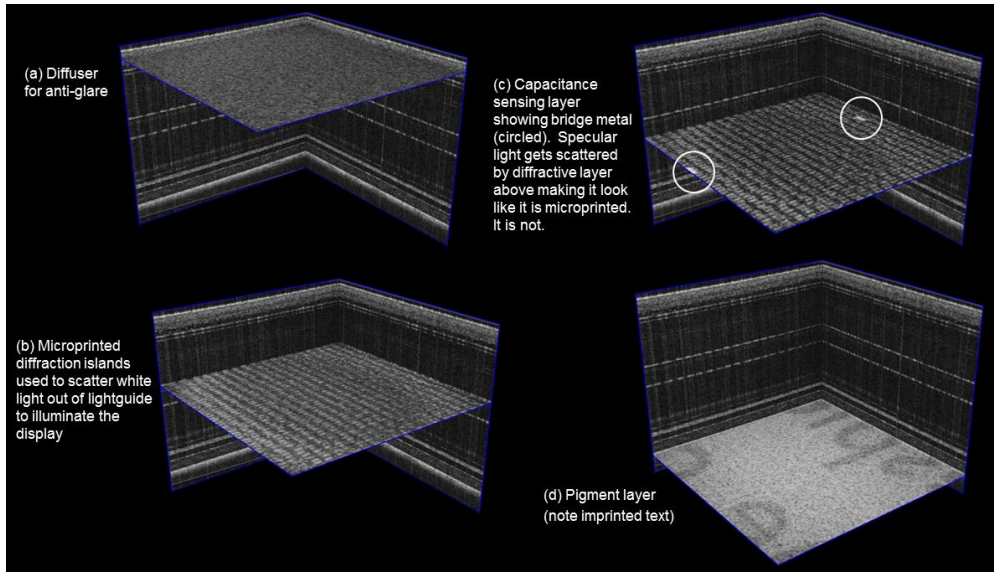

Figure 3. Three dimensional renderings of OCT data showing the (a) diffuser, (b) scattering layer, (c) touch-screen capacitance layer, (d) E Ink pigment layer.

## 3. Light diffuser

From close examination of Figure 2, the light diffuser appears to be four thermoplastic layers containing small air bubbles. Microscope photos in Figure 4 show one image focused on the diffuser surface and another focused on the pigment layer below. The first image shows the diffuser bubbles and the second shows individual pixels in the pigment layer, slightly blurred because of the diffuser. The pixel pitch is 120 microns. We speculate that the diffuser material is polymethylmethacrylate (PMMA), which can be formulated with varying levels of bubble content.

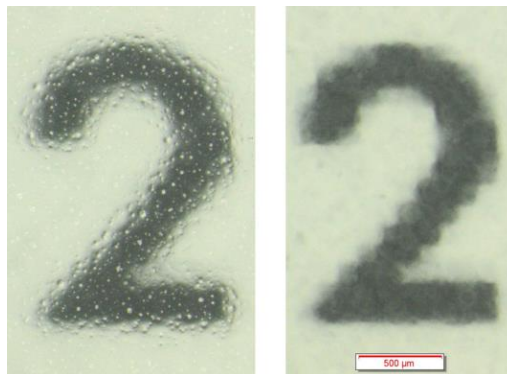

Figure 4. Microscope photo focused on the light diffuser surface (left) and on the E Ink pigment layer (right).

## 4. Lightguide illumination layers

The internal white light illumination is a new feature in the Paperwhite model. Four white LEDs shine into a planar waveguide and microprinted scattering centers are designed to uniformly scatter light out of the guide onto the pigment layer. The success of this scheme is shown in the illumination uniformity map of Figure 5. The white/black contrast of the electrophoretic display is 17:1. These false color images were obtained by photographing the display using a RAW file format and removing the gamma compression from the final bitmap images to register linear light intensity.

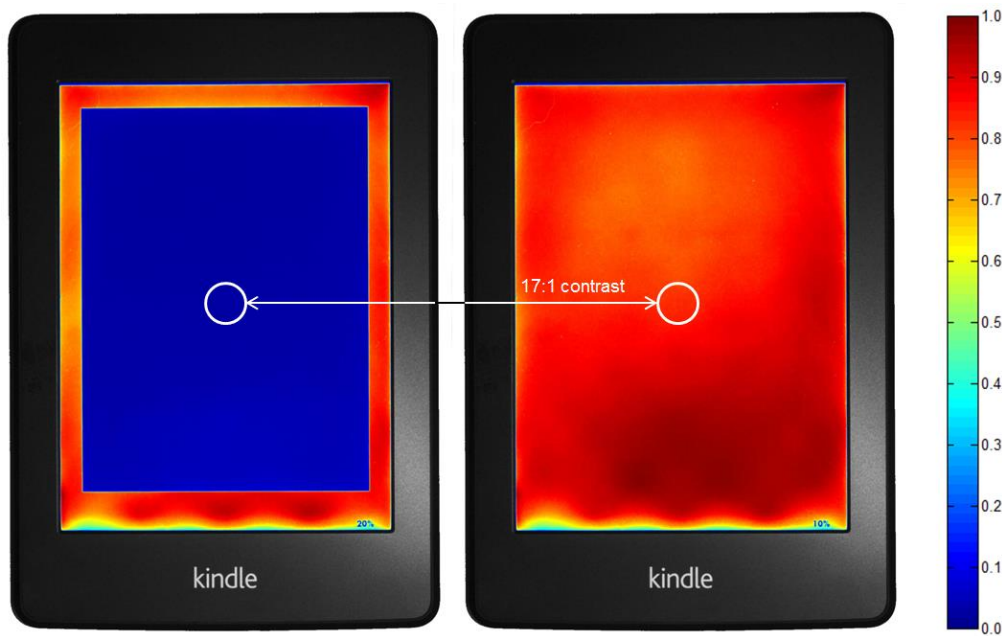

Figure 5. False color light intensity maps for a black page with white border (left) and white page (right). Four white LEDs inject light into the lightgide at the bottom of the display and it propagates towards the top, being scattered onto the electrophoretic display along the way.

The scattering centers are apparently imprinted on the surface of the high-index guide layer as described in these patents [3,4]. Each scattering center is a small island corrugated plastic as shown in Figure 6. The islands are smaller at the bottom of the display, near the white LEDs. For uniform illumination, the scattering needs to increase near the top of the display as the white light is depleted. These scattering islands are seen in the OCT image in Figure 3b as a kind of row-like structure in the coarser pitch dimension. These scattering centers are visible on a disassembled Paperwhite near the edge of the display where the centers overlay the metal film on the capacitance layer. The pitch of the centers is roughly 200 x 100 microns, although there appears to be some randomness of placement and of corrugation angle. Figure 7 shows the increased island size at the top of the display, away from the four illumination LEDs at the bottom.

We speculate that the lightguide material is polycarbonate (PC), which is capable of being microprinted, as is done in the case of pressed compact discs, for example. OCT measurements of the optical and physical thickness of these layers indicates the refractive index is greater than 1.5, consistent with published numbers for PC at a 1 micron wavelength. This high index would provide light guiding when sandwiched between lower index materials.

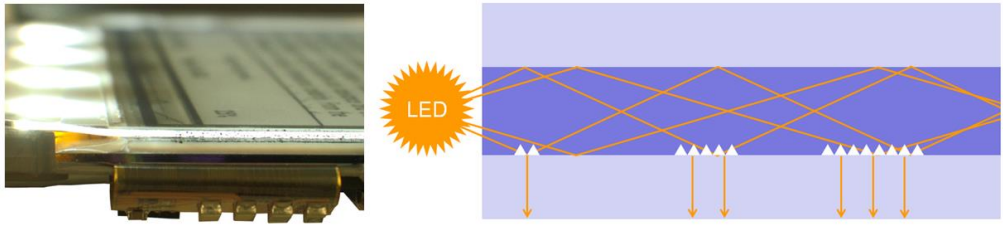

Figure 6. Microscope photograph of light emitted from the edge of the lightguide of a disassembled Paperwhite with the four illumination LEDs out of focus in the background (left). Structure of the lightguide and scattering centers (right).

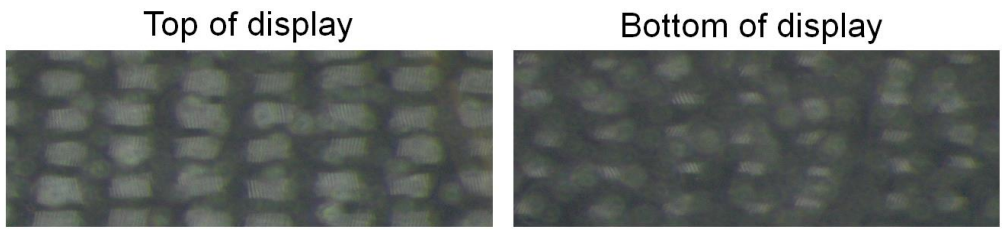

Figure 7. Microscope photographs of the microprinted scattering centers. The image is not clear because of the intervening diffuser layer. Close examination shows the corrugations from the microprinting.

**5. Touch screen capacitor layer**

The capacitance layer shown in Figure 3c appears to be corrugated like the lightguide, but that is an artifact of the OCT imaging. The display was tilted in the OCT experiment to eliminate strong specular reflections. The result is that the specular beam from the capacitance layer misses the detector, but will get to the detector by scattering in the layer above as drawn in Figure 8. This modulates scattering islands onto the capacitance layer signal giving the mistaken impression that the capacitance layer is corrugated as well.

The reflections from small metal traces circled in Figure 3c are real. Those are apparently metal crossover traces connecting one dimension of the touch screen projected capacitance grid. They are imperceptible to a person reading an E-book, but can be seen under a microscope. We were unable to image the capacitor traces, likely made from indium tin oxide, with standard OCT and even phase sensitive OCT. The Paperwhite has a multi-touch capability and we are presuming that the capacitance layout is similar to that drawn in Figure 9.

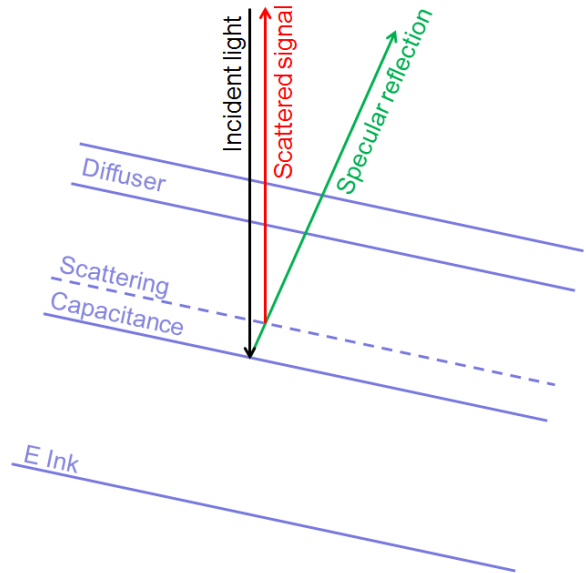

Figure 8. Optical path from the capacitance layer to the OCT detector, showing how the scattering layer affects the signal and confuses the interpretation of the images.

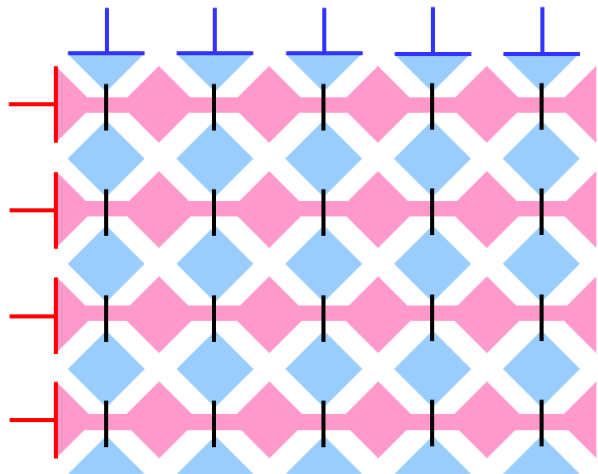

Figure 9. Diamond-shaped transparent pads in a typical projected capacitance array used for multi-touch sensing. The crossover traces imaged by our OCT system are shown in black. The dark blue and dark red metal traces that connect to the presumed ITO pads can be seen under a microscope and by OCT.

## 6. E Ink electrophoretic display layers

OCT can image characters formed on the E Ink electrophoretic display layer, although the contrast at 1060 nm is not high. This is shown in Figure 10 in three dimensions along with reflections of all the intervening layers of the display.

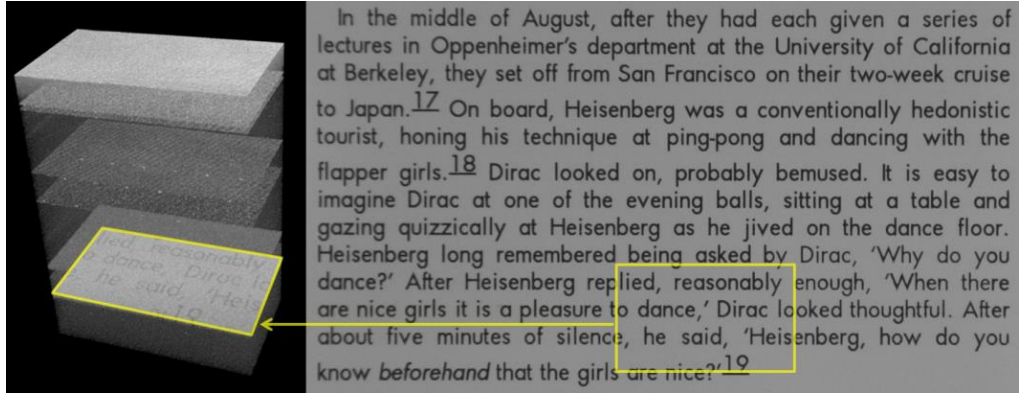

Figure 10. Three dimensional rendering of OCT data (left) and photograph (right) of the Kindle Paperwhite display with text from [8] displayed.

While the display state, white/black/gray, is static with the power removed, the pigment particles exhibit Brownian motion. The particles are held in a semi-permanent potential well, but they are still free to move somewhat within the liquid. This can be seen from the speckle pattern in the M-mode image of Figure 11 that shows 250 A-lines from a stationary beam on the Kindle display. All the display layers and interfaces are seen, with unchanging speckle patterns except for the E Ink pigment layer. The light diffuser, for example, shows stable horizontal speckle lines over the 5 minute measurement period. The E Ink pigment speckle pattern changes for each A-line, showing that the pigment particles are in motion. This particular measurement was taken from a Kindle Paperwhite display in the white state whose battery was removed some months before.

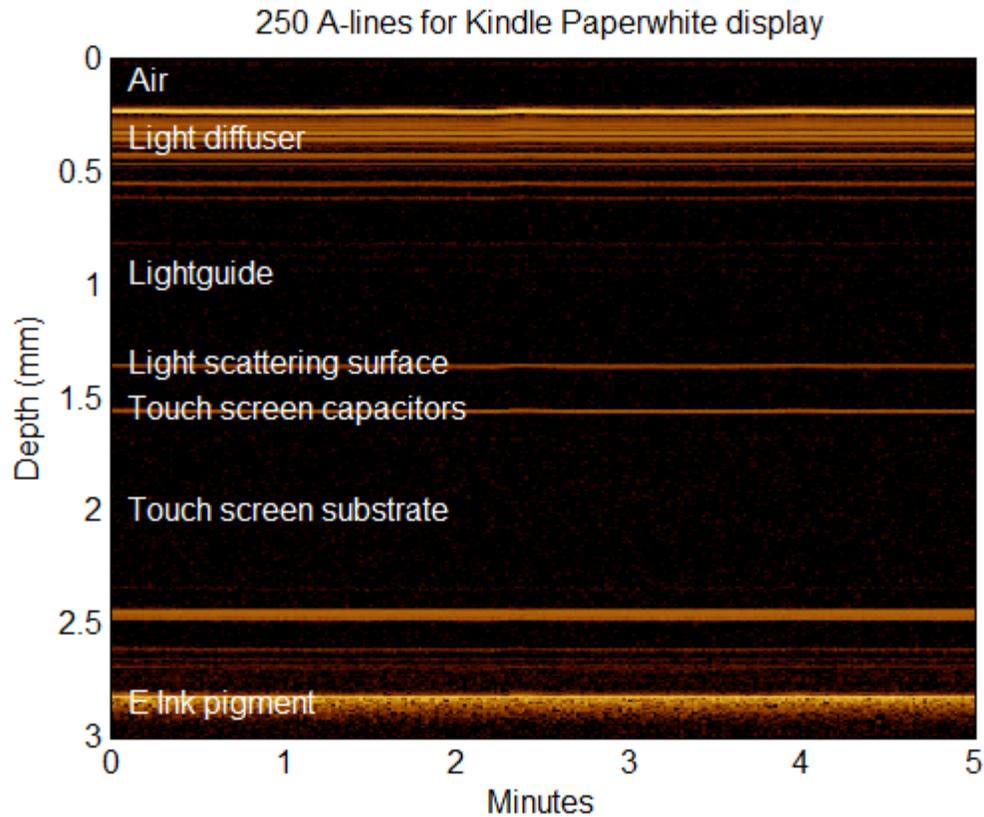

Figure 11. M-mode image formed from 250 A-lines taken over a 5 minute time period. Stable speckle patterns are seen everywhere except within the E Ink pigment layer, showing that the pigment particles, while trapped in a stable potential well, still exhibit Brownian motion.

An electrophoretic display has fluid-filled microcapsules containing mobile, charged pigment particles that can be driven towards or away from the viewer by electrophoretic forces depending on the direction of an electric field across the capsule layer. The Kindle Paperwhite has black and white particles. The motion of the white particles can be tracked by phase sensitive OCT since they are highly reflecting. Movement towards or away from the display surface can be detected, but lateral motion cannot since it produces no Doppler shift. Therefore motion due to electrophoresis can be measured, but motion from dielectrophoretic forces [9] cannot.

The Kindle Paperwhite was mounted under a stationary beam from a 1060 nm, 100 kHz swept source [5]. A 12-bit data acquisition board [5] was used to stream 8.2 seconds of data to computer memory over a PCIe interface. While the data converters are capable of 550 MS/s rates, the board was actually clocked from a k-clock interferometer at frequencies ranging from 170 to 330 MHz. The fiber-based Mach-Zehnder k-clock drifts with temperature and the starting wavelength of the laser sweep jitters a few clock pulses sweep to sweep. These are not good conditions for a phase sensitive measurement and $2\pi$ phase errors can easily be made. We use a new phase unwrapping algorithm [10] that is tolerant to phase jitter to combat that problem. There are both one and two dimensional versions of the method.

Figure 12 shows the results of an experiment where the Kindle Paperwhite display started white, was switched to black, and then back to white. Doing that required swiping the touch-screen display with a finger to "turn the page." Physically touching the display moved it by about a dozen microns, even though the unit was firmly strapped down. Phase data from both the display surface and the pigment layer were collected so the overall device motion could be subtracted to just obtain the pigment motion. This creates a "virtual" common path interferometer, which is more phase stable. This worked out fairly well, as shown in Figure 12. The red curve shows about 4 microns of white particle motion away from the display surface when it is switched to the black state. This is in the face of about 12 microns of overall motion from a finger pushing the display away from the OCT probe. A few $2\pi$ phase errors, which amount to about 0.5 microns, are likely, but overall the measurement looks reasonable.

The Kindle Paperwhite is sluggish, taking about 0.5 seconds to respond to a finger command. The display reflectivity changes in synchronism with the pigment movement, not the finger motion. The 1060 nm laser signal from the pigment layer suffers from speckle effects and is very "noisy". A simultaneous measurement of white light reflectivity shows a much cleaner trace. However, both measurements track and are synchronized with the pigment motion.

It is clear that the white particles move away from the surface when the display transitions to the dark state, and the data shows this. The registered motion is only 3 microns, although the E Ink microcapsules are believed to be much bigger. Quantifying the movement in microns is problematic since there will be some signal from the stationary microcapsule walls and from the black particles that move in the opposite direction. It is expected that the signal will be dominated by white particle reflection only when they are near the top surface of the microcapsules. A more sophisticated model of the optical interaction is needed to probe further.

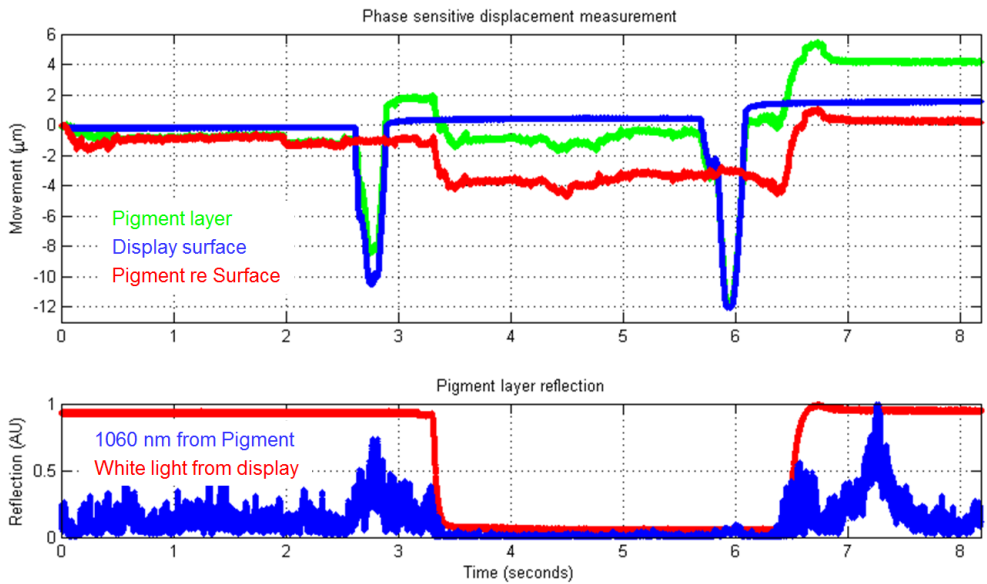

Figure 12. Phase sensitive displacement measurement of pigment particles (top) and display reflectivity (bottom).

## 7. Thin film transistors

A thin-film transistor (TFT) array drives the pixels in the electrophoretic display. The TFTs are not accessible optically from the front of the display since the pigment layer prevents light from penetrating that far. The TFTs can be imaged from the back of the display, nondestructively, by disassembling the Paperwhite. The TFTs are fabricated on a glass substrate and the layers deposited on the substrate can be imaged, although most of the deposited materials are metal and light does not penetrate further.

Figure 13 shows three images of a 3x3 cell portion of the TFT pixel driver array. The cell pitch is 120 microns. The OCT images are of limited use because of the poor lateral resolution, however there is potential for added diagnostic information with a higher resolution scanner, especially with phase sensitive imaging. The phase sensitive image was made by subtracting the TFT layer phase from the substrate phase and applying the two-dimensional version of the filtered phase unwrapping algorithm of Ref. [10].

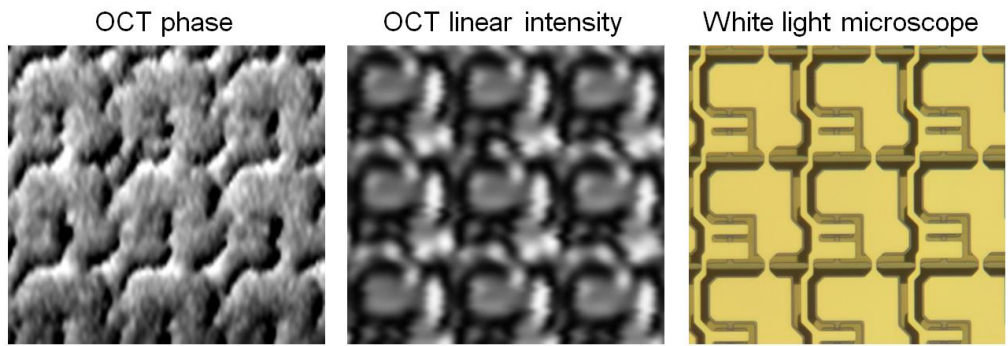

Figure 13. 3x3 cells of the thin film transistor array seen with phase sensitive OCT (left), standard OCT (center), and white light microscopy (right)

## 8. Summary

This work illustrates an industrial application of OCT and reveals the impressive electro-optical technology behind the Kindle Paperwhite E-book reader. These measurements were made nondestructively. The reader was taken apart, but still worked when reassembled. An advanced 1060 nm swept source and a new data acquisition board capable of streaming large data sets made this work possible. These experiments also demonstrate a new algorithm for phase-unwrapping [10] that makes phase sensitive measurements possible in the face of modest laser phase jitter.

**References and links**